\documentstyle[12pt]{article}
\begin{document}
\title{ CORRELATIONS AND FLUCTUATIONS '98   }
\author{A. BIALAS\\ 
 M.Smoluchowski Institute of Physics, Jagellonian
University \\ Reymonta 4, 30-059 Krakow, Poland \\ E-mail:
bialas@thp1.if.uj.edu.pl} 
\maketitle
\begin{abstract}
Summary talk at the VIIIth International Workshop on Multiparticle
Production "Correlations and Fluctuations '98" held at Matrahaza, Hungary from
14th till 21st of June 1998.
\end{abstract}

\section{ Introduction}

To begin my summary, let me first single out two contributions to the
meeting which I consider as the real steps forward in our quest for
understanding the complicated phenomena of multiple production, but
which I am unable to summarize in short terms.

 First, Hans Eggers showed
an amazingly simple and elegant solution of the model of multiplicative
cascade \cite{eg}. As we all know, the model played an important role in
formulation of  the subject of this meeting, i.e. studies of
fluctuations in multiparticle production. It seems to me very likely that this
new development will soon create further progress in our field.

 The second
contribution  is that by Bo Andersson \cite{an2}. In the
investigation of  the structure of the QCD cascade at the end
of the available phase-space, the Lund group  arrived recently at the conclusion
that emitted gluons are ordered not only in rapidity but
also in azimuthal angle  and - moreover - proposed the method to
measure this effect. If this is indeed confirmed by experiment, this result
would mean a significant step forward in understanding the QCD cascade which is
the major problem in description of the multiparticle production processes.

\section {Bose-Einstein interferrence}

Coming to the bulk of the conference, it was clear to everybody that
this year the discussion of the Bose-Einstein interference or, in other
words, the Hanbury-Brown and Twiss effect was the dominating issue. Let
me thus start by a brief reminder what is this all about\footnote{The
{\it physics} of the HBT effect was recently extensively 
reviewed by G.Baym \cite{by}}.

The practical problem we face can be formulated as follows: given 
  a calculation (or a model) which
ignores identity of particles,  how to "correct" it 
in order to take into account the effects  quantum interference (which is the
consequence 
 of the identity\footnote{ It should be understood that this
problem is very common in quantum mechanical calculations, as
illustrated, e.g., by eveluation of Feynman diagrams. I would like to
thank J.Pisut and K.Zalewski for discussions of this question. }). Let us
thus suppose that we have an amplitude for production of N particles
$M^{(0)}_N(q)$ ($q=q_1,...q_N$) calculated with the identity of
particles being ignored. The rules of quantum mechanics tell us that, to
take the identity of particles into account, we have to replace
$M^{(0)}_N(q)$ by a new amplitude $M_N(q)$ which is a sum over all
permutations of the momenta $(q_1,...q_N)$
\begin{equation}
M^{(0)}_N(q) \rightarrow  M_N(q) \equiv \sum_P M^{(0)}_N(q_P)   \label{1}
\end{equation}
This would be the end of the story if particle production was described by a single
matrix element. In general, however, we have to average over parameters which
are not measured and therefore the correct description of the multiparticle
final state is achieved in terms of the
density matrix
\begin{equation}
 \rho^{(0)}_N(q,q') =\sum_{\omega} M^{(0)}_N(q,\omega)
M^{(0)*}_N(q',\omega),            \label{2}
\end{equation}
rather than in terms of a single production amplitude. The sum in
(\ref{2}) runs over all quantum numbers $\omega$ which are not measured
in a given situation. $\rho^{(0)}(q,q')$ gives all available
information about the system in question. At this point it is useful to
note that, when tranformed into (mathematically equivalent) Wigner
representation
\begin{equation}
W_N(\bar{q},x) = \int d(\Delta q) e^{ix\Delta q} \rho^{(0)}_N(\bar{q},\Delta q)
\label{3}
\end{equation}
($\bar{q}= (q+q')/2; \;\; \Delta q =q-q'$)  it gives information about the
distribution of momenta and positions
of the particles (see, e.g., \cite{bk} for a discussion of this point).

 Using (\ref{1}) one
easily arrives at the formula for the {\it corrected}
 (i.e., with identity of particles taken into account)
density matrix $\rho_N(q,q')$
  and one finally obtains the observed multiparticle density
\begin{equation}
\Omega_N(q)= \frac1{N!} \sum_{P,P'} \rho^{(0)}(q_P,q_{P'})   \label{4}
\end{equation}
where the sum runs over all permulations $P$ and $P'$ of the momenta
$(q_1,...q_N)$. The factor $\frac1{N!}$ appears because the phase space for $N$
identical particles is $N!$ times smaller than the phase space for $N$
non-identical particles. The formula (\ref{4}) is in common
use\footnote{Using the hermiticity property  of the density matrix, the
double sum in (\ref{4}) can be reduced to a single sum. The factor
$\frac1{N!}$ is then absent.}  and is the basis of our further discussion.

\subsection {A theoretical laboratory: independent particle production}

The case of independent particle production is an attractive theoretical
laboratory which, although not expected to describe all details of the data,
reveals -nevertheless- some general (and generic)  features of the problem.
This was first recognized by Pratt \cite{pr}.
In terms of the density matrix, the independent production means that
the density matrix factorizes into a product of single-particle density
matrices
\begin{equation}
\rho^{(0)}_N(q,q') =
\rho^{(0)}(q_1,q_1')\rho^{(0)}(q_2,q_2')....\rho^{(0)}(q_N,q_N')
\label{5}
\end{equation}
and that the multiplicity distribution is the Poisson one
\begin{equation}
P^{(0)}(N) = e^{-\nu}\frac{\nu^N}{N!} .        \label{6}
\end{equation}

Several contributions to this problem were presented at the meeting
\cite{zi,zh,si,su}. It turns out \cite{zi,zh} that in the case of a
Gaussian density matrix  the problem can be solved analytically.
             The main results (valid
also in  the general case of an arbitrary density matrix \cite{bz}) can be
listed as follows.

(a) All correlation functions $K_p(q_1,...,q_p)$ and the single particle
 distribution $\Omega(q)$ can be
expressed in terms of one (hermitian)  function $L(q,q')=L^*(q',q)$
 of two momenta:
\begin{eqnarray}
\Omega(q)= L(q,q); \;\; K_2(q_1,q_2)=L(q_1,q_2)L(q_2,q_1)\nonumber \\
K_3(q_1,q_2,q_3)= L(q_1,q_2) L(q_2,q_3)L(q_3,q_1)+ L(q_1,q_3) L(q_3,q_2)
L(q_2,q_1), \label{7}
\end{eqnarray}
and analogous formulae for higher correlation functions.

(b) At very large phase-space density, particle distribution approaches a
singular
point representing the phenomenon of Bose-Einstein condensation: almost all
particles populate the eigenstate of $\rho^{(0)}(q,q')$ corresponding to the
{\it largest eigenvalue}. The resulting multiplicity distirbution is very broad
(almost flat) so that, e.g., probability of an event with no single $\pi^0$
produced is non-negligible. Such a situation may perhaps be a possible
explanation of the somewhat elusive "centauro" events \cite{fu}, as suggested
by Pratt \cite{pr}\footnote{This effect was also considered in connection
with the possible production of the Disoriented Chiral Condensate \cite{blk,bj}.
The present argument adds another obstacle on the difficult road to
observation of DCC, as discussed thoroughly at this meeting by the Bergen group
\cite{cse}.}.

(c) At high density, the parameters extracted from the observed spectra have
little in common with the input parameters characterizing the source. In
particular, for the Gaussian source in  the BE condensation limit we have
\begin{equation}
R_{eff}^2 =\frac {R^2}{2R\Delta}<R^2;\;\;\;\Delta_{eff}^2 =\frac
{\Delta^2}{2R\Delta} <\Delta^2       \label{8}
\end{equation}
where $R^2=<x^2>$  and $\Delta^2=<q^2>$ are the average values of the position
and momentum of the particles (uncertainty condition implies $R\Delta \geq
1/2$.)

It should be not surprizing that the very restrictive condition of
independent production, as expressed by (\ref{5},\ref{6}), is not
realized in nature. This was shown at the present meeting by Lorstad
\cite{lo}, who demonstrated that there are practically no genuine
three-particle correlations\footnote{The importance of the absence of
3-particle correlations in heavy ion collisions was emphasized already
some time ago  \cite{is}.} in S-Pb collisions at CERN SPS. Since the
two-particle correlations are clearly visible, this observation cannot
be reconciled with Eq.(\ref{7}). It was also shown by Eggers et al \cite{elb}
that the UA1 data are in contradiction with (\ref{7}), although in this
case the 3-body correlations seem to be too large to satisfy (\ref{7}).
This striking difference between the behaviour of heavy ion and
"elementary" collisions is certainly very interesting and deserves
further attention.

We cannot thus consider the results obtained from (\ref{5},\ref{6}) to be
realistic description of the data. Nevertheless, the main conclusion about the
possibility of Bose-Einstein condensation remains an interesting option which
is worth serious consideration.

\subsection { Monte Carlo simulations}

In this situation, the practical method to study the effects of BE
symmetrization on particle spectra is to implement it into the Monte
Carlo codes. A "minimal" method of performing this task was suggested
some time ago  \cite{bk}. The idea is to take an existing code (which
reproduces the distribution of particle momenta, i.e. the diagonal
elements of the density matrix) and to modify only the off-diagonal
elements of the multiparticle density matrix (\ref{2})\footnote{As seen
from (\ref{3}) this corresponds to introducing an - a priori arbitrary -
distribution of particle emision points in configuration space.}. Each
event generated by the MC code is then given a weight which is
calculated as the ratio of symmetrized distribution [Eq.(\ref{4})], and
the unsymmetrized one. In this way the modification of the original
spectra is kept at the minimum.

A  practical realization of this idea has been developped by the Cracow group 
 \cite{fw} and was presented by Fialkowski at this meeting.
 They  propose the unsymmetrized density matrix in the form
\begin{equation}
\rho^{(0)}_N(q,q')= P_N(\bar{q}) \prod_{i=1}^N w(q_i-q_i')   \label{9}
\end{equation}
where $P_N(q)$ is the probability of a given configuration obtained in
JETSET and $w$ is a Gaussian. 
This prescription does not modify the diagonal elements of the
unsymmetrized density matrix ($w(0)=1$) and, moreover, does not
introduce any new correlations between emission points of the produced
particles (when transformed into Wigner representation, the product
$\prod w(q_i-q_i')$ becomes the product $\prod w(x_i)$). Thus (\ref{9})
can indeed be considered as a minimal modification of the existing code.
The autors find that this prescription represents well the existing data
on two-particle correlations and that they can recover the experimental
multiplicity distribution by a simple rescaling with the formula $P(N)
\rightarrow P(N)cV^N$, without the necessity of refitting the JETSET
parameters.

The results presented at this meeting concerned the $W$ production in
$e^+e^-$ collisions. The authors find that the expected mass shift is very
small (less than 20 MeV). They also predict a shift of multiplicity
observed in hadronic decay of one and two $W$'s: 
\begin{equation}
 n(2W)-2 n(W) = 2.1 \pm 0.9 \label{10} 
\end{equation}
 This may be an
overestimate because (as seen from (\ref{9})), in present version of the
model the position of particle emission point is not correlated with its
momentum, whereas this effect is likely to be present in reality.

A more fundamental approach has been pursued  since some time by
Andersson and Ringner \cite{ar}. 
It is based on the famous paper by Andersson and
Hoffman \cite{ah} and was presented here by Ringner and by
Todorova-Nova \cite{tn}. They write the
"uncorrected" matrix element for the deacay of one Lund string in the form
\begin{equation}
M^{(0)}_N (q) \sim exp\left[\left(\frac{b}2+\frac{i}{2\kappa}\right)A(q)\right]
\prod_{i=1}^N e^{-\frac{\pi}{2\kappa} q_{\perp i}^2}  \label{11}
\end{equation}
and then follow the procedure explained in introduction. Two particle
correlations are well described and several interesting effects are
predicted. Among them: (a) the longitudinal and transverse correlations
are expected to be different because they are controlled by two
different physical mechanisms; (b) Three particle correlations are
predicted non-vanishing and were actually calculated; (c) $WW$
production was studied and no significant mass shift is expected; (d) No
multiplicity shift in the $W$ decay is predicted.

This last conclusion is a consequence of the fact that, in case of more
then one string present in the final state, no symmetrization between
particles stemming from different strings is performed. This corresponds
to the assumption that the strings are created at a very large distance
from each other. One thus may expect that in a more realistic treatment
some multiplicity shift should be present\footnote{In both \cite{fw} and
\cite{ar}  the "interconnection effect" \cite{el} (which has tendency to reduce the
multiplicity) is neglcted. The full phenomenological analysis of the
data is therefore certainly more complicated.} .

\vspace{6.5cm}

Fig.1. The second order cumulant plotted versus inverse of the rapidity
density. Data from UA1 \cite{bu}.

The problem of quantum interfererence between particles from different
strings is certainly the important one and its solution may be crucial
for the success of the Lund model in processes which are more
complicated than $e^+e^-$ annihilation. In this context interesting data
of UA1 collaboration were presented By B.Buschbeck \cite{bu}. The
authors studied the dependence of the correlations between like- and
unlike- pairs as function of the particle density. The data are shown in
Fig.1. One sees linear dependence of the normalized cumulants on
$\frac1{dN/dy}$. One observes, furthermore, that in the region $Q=7$GeV
(where no HBT effect is expected) the cumulant vanishes in the limit of
large density. On the other hand, in the region $Q=.1$GeV (which is
likely to be dominated by BE correlations) the cumulant tends to a
finite value in this limit.

 To understand the meaning of these data,
consider  particle
emission from a number $N$ of {\it independent} sources.  In this case the
particle density is
\begin{equation}
\frac{dn}{dy} = N \frac{d\nu}{dy}    \label{bu1}
\end{equation}
where $\frac{d\nu}{dy}$ is the particle density from one source. The
normalized two-particle correlation function is
\begin{equation}
\frac{\frac{d^2n}{dydy'}}{\frac{dn}{dy}\frac{dn}{dy'}}|_{y=y'} -1 =
\frac1{\frac{dn}{dy}} K(y)   \label{bu2}
\end{equation}
where K(y) depends only on the particle distribution from one source.

Thus the emission from several independent sources implies that the
normalized cumulant is {\it inversely proportional} to the particle
density. As seen from Fig.1, this is in good agreement with the data at
large $Q$. At small $Q^2$, however, dominated by BE correlations, the
normalized correlation function approaches a constant different from
zero at large densities, in disagreement with (\ref{bu2}). This implies
correlations between the sources which may well originate from the
quantum interference between them. The qualitatively different behaviour
in the two regions supports this idea. On the other hand, the fact the
 that correlations between the like- and unlike charges behave similarly (see Fig.1)
casts a doubt on this interpretation. Further investigations along these
lines are thus certainly needed.

Finally, let me comment on the contribution of Lonnblad \cite{lon}. 
He presented the method of describing the HBT effect by shifting the
momenta of 
identical particles so that the observed two-particle correlations are
reproduced. 
Personally, I do not believe that this is a correct way of treating
the problem of BE interference\footnote{It   seems to be an attempt to
treat the effects of quantum interference as a final state
interaction.}. I like thus  only mention the contribution to this meeting
 by Smirnova \cite{sm}, who showed that the method presented by
Lonnblad does not reproduce the values of the parameters used as an
input.

\subsection { Probing the space-time structure}

Much attention during the meeting was devoted to the information one may
obtain from the data on quantum interference about the space-time
structure of the multiparticle system created in the collision. Although
such analyses can have at most a limited scope, as they only provide
the information about the system at the freeze-out and, as emphasized by
Weiner \cite{we}, require several additional assumptions - they provide
nevertheless a unique opportunity to investigate this problem. Most of
the {\it caveats} are thus usually postponed to the future (and better
data) and the analysis is carried on.

The presented investigations were based on the hydrodynamic approach. The
general framework was explained by Csorgo \cite{cs2} who advocated a new
Buda-Lund parametrization, as an improvement with respect to the
standard YKP one \cite{ya}. There were four presentations of the
experimental results.

 Lorstad discussed the $m_t$ dependence of the data of NA44 and LEP
\cite{lo}. The radius of the system systematically decreases with
increasing transverse mass of the particles. In case of heavy ion
collisions this  is usually interpreted as evidence for hydrodynamic flow.
However, the same phenomenon is observed also in $e^+e^-$ annihilation
where the notion of hydrodynamic flow is perhaps not so easy to accept.

 Seyboth presented data of NA49 experiment on Pb-Pb collisions at SPS
\cite{se}. He showed rather convincingly that (i) The longitudinal flow
of particles is well consistent with the Bjorken in-out model \cite{bj2}
and (ii) Particle emission starts rather late and it lasts not very long:
life time of the system is of the order of 8 fm, while the duration of
pion emission is only of the order of 3 fm.

\vspace{6.0cm}

Fig.2. Space-time region of particle emission in S-Pb collisions
\cite{st}.

These two features are also present in the NA44 data on S-Pb collisions,
discussed by Ster \cite{st}. This is seen in Fig.2 where the
reconstructed space-time distribution of the source of particles is
shown. One clearly sees a characteristic Bjorken shape of the source.
One also sees that particle emission in the central region starts only
at about 4 fm and is practically finished $\sim$1.5 fm later.

A qualitatively similar behaviour is found in $\pi p$ collisions of NA22
experiment, as presented by Hakobyan \cite{ha}. The picture shown in
Fig. 3 looks qualitatively rather similar to that in Fig.2.  Note,
however, an important {\it quantitative} difference: In hadron-hadron
reaction the particle emission in the central region starts almost
immediately after collision and lasts about 1.5 fm.

\vspace{11.0cm}

Fig.3. Space-time region of particle emission in $\pi$-p collisions
\cite{ha}.

We have also seen from a contribution of Schlei \cite{sh} 
 that the BE correlations may serve as a tool for analysis (and
improvement) of the equation of state of the strongly interacting
matter. The reason is that the volume ocupied by the system at
freeze-out depends on the equation of state and thus information on this
volume provided by BE correlations restricts severely the possible
equations of state. For esentially the same reason information from BE
correlations helps to estimate the particle density in phase-space, as
was pointed out by Pratt \cite{pr3}.

This completes the list of contributions discussing the data on BE
interference. Several other results about particle correlations related
to the space-time structure were also shown.

Lednicky \cite{le} presented an interesting idea that correlations
between the {\it non-identical} particles can provide information on the
time sequence of their production. Indeed, consider two particles moving
in the same direction and suppose that they are subject to a final state
interaction (for instance Coulomb interaction). If the faster one is
emitted before the slower one, the effect of the interaction shall be
smaller (because they move apart from each other), otherwise it will
be stronger (because the faster particle will catch the other one). The
feasibility of such measurements was discussed and the prospects seem to
be promising.

Kuvshinov \cite{ku} discussed production of instantons in deep inelastic
collisions. The most striking effect seems to be a very narrow
multiplicity distribution in the instanton decay, which may serve as a
good signal of such a phenomenon.

Particle-antiparticle correlations were discussed by Andreev \cite{and}
and by Csorgo \cite{cs3}. They considered a passage of a particle
through a region of the false vacuum (DCC) if such a region was
produced in a collision. Since the particle in the false vacuum cannot
be on its mass shell, an adjustment of the wave function is necessary
when it leaves the DCC region. This  manifests itself as additional
particle production. Since the quantum numbers of the produced system
must be those of vacuum, one concludes that a particle-antiparticle pair
with opposite momenta must show up. This would be certainly a very
attractive signal for the DCC. The estimates of Csorgo et al \cite{cs3}
are that the effect is expected to be fairly strong (although precise
estimates are not possible at the moment) and thus deserves attention of
experimenters.

Several papers on possible evolution of DCC were presented by the Bergen
group \cite{cse}. Unfortunately I have neither space nor competence to
comment on them.

\section {Multiplicity distributions}

Multiplicity distributions were discussed in several contributions.

Giovannini and Ugoccioni \cite{gi} presented estimates of the KNO
scaling violation expected because of the onset (and eventually
dominance) of hard scattering (mini-jets and jets) at high energies.
Hegyi \cite{heg} discussed the generalized negative binomial
distribution along the lines proposed some time ago by Carruthers, and
also suggested some interesting ideas about the scaling violation.

Ploszajczak \cite{pl} discussed scaling laws in
the systems which undergo the 2nd order phase transition. The numerical
analysis leads to the conclusion that such systems obey the scaling law
of the form 
\begin{equation} 
<n>^{\delta} P(n) = \Phi(z_{\delta})
\label{12}
 \end{equation}
 with 
\begin{equation} 
z_{\delta}=\frac{n-<n>}{<n>^{\delta}} \label{13}
 \end{equation}
where $\delta$ is a number between 1 and $\frac12$. It turns out,
furthermore, that the value of $\delta$ is determined by the nature of
the variable $n$. If the system is at the phase transition and $n$ is
the correct order parameter, then $\delta =1$. If the system is at phase
transition but $n$ is not the  correct order parameter, $\frac12 <\delta <1$.
What is most unexpected, however, even if the system is {\it not} at
phase transition one still has the scaling law (\ref{12}) with $\delta =
\frac12$. This intriguing property certainly requires (and deserves!)
further investigation.

\section {Perturbative  and non-perturbative QCD}

Application of perturbative QCD calculations to multiparticle spectra
was a subject of a hot discussion. It is now rather well established
that the average multiplicity and single particle spectra are well
described by perturbative QCD supplemented with the principle of
parton-hadron duality \cite{phd}. At this meeting Lupia \cite{lup}
presented a calculation of the cumulants of the multiplicity
distribution and showed that they also agree with the data. Thus the
principle of parton-hadron duality is now extended even for integrated
correlation functions. This statement was challenged by Mangeol
\cite{man} who analyzed the data on cumulants in jets and found that the
results do not obey the predictions of perturbative QCD in the region
where one expects them to be the best, i.e. in high energy jets. The
problem requires certainly further discussion but it is clear that the
Lupia calculation marks an important step towards understanding the
meaning of perturbative QCD predictions and of parton-hadron duality.

The real challenge to the idea of parton-hadron duality is to explain
the data on diffrential correlaton functions. Indeed, it is hard to
understand how the momenta of the produced hadrons can follow so closely
the momenta of the created partons that the correlations between them
are not washed out\footnote{This problem does not arize, of course, if
one considers only the integrated correlation functions.}. Therefore a
non-trivial extension of the principle of parton-hadron duality must be
formulated in order to give quantitative meaning to perturbalive
calculations of multiple production. This point of view was
substantiated by Kittel \cite{ki} who showed that the predictions of
perturbative QCD formulated some time ago \cite{pqcd}, are badly
violated by the L3 data. On the other hand, the same data are well
described by the JETSET code. The conclusion is that the hadronization
part is probably not correctly taken into account by the simple (naive?)
parton-hadron duality. This conclusion was challenged by Ochs \cite{oc}
who showed in his talk that the previously published calculations
included several simplifying assumptions (the most important among them
seems the neglect of energy-momentum consevation) and thus it is not
clear which part of the result is actually responsible for the
failure. Ochs presented several improvements and indicated the
kinematical regions where perturbative QCD effects have a better chance
to be seen. In my opinion, further work on these lines is necessary,
however, to establish a reliable, quantitative link between partons and
hadrons and to determine its range of application.

Another philosophy was presented by Hwa \cite{hw} who considered a
fundamentally non-perturbative approach to the problem of multiparticle
production. The model is an implementation of an old idea of Feynman
\cite{fe} in which the partons in the final state are just those present
already in the initial state but rearranged during the collision. Hwa
supplements this idea with a specific prescription
for the transition from partonso hadrons:
  the neutralization of colour happens by
random walk in colour space. According to the author, the main
advantage of this mechanism is that it can provide a natural explanation
for the fractal character of multiparticle spectra \cite{bp} which is
now firmly established in $e^+e^-$ and hadron-hadron data (see, e.g.
\cite{ddk}). Further work is needed, however, to confront the details of
the model with experiment.

Another interesting contribution was presented by Chekanov \cite{che}
who discussed the forward-backward correlations in deep inelastic $ep$
scattering. He showed that the perturbative QCD calculations predict
negative correlation between multiplicities of the current and target jet
(in the Breit system). The data appear to follow this prediction (within
fairly large errors). This seems to be an important step in more precise
definition of the parton-hadron duality, indicating that it works not
only for fully integrated quantities but also for those integrated over
a large enough regions of phase-space.

\section {"Traditional" intermittency}

"Traditional" intermittency analysis of the data was presented by
Sarkisyan \cite{sar} representing OPAL experiment. He showed that
 these data cannot be fully explained
by the MC codes (JETSET and HERWIG were used) at very small phase-space
intervals. The data in three-dimensional bins show a rather clean
power-law behaviour and it will certainly be very interesting to see the
results of the fit determining the intermittency parameters. It shall be
also interesting to see if inclusion of BE correlations into the MC
codes can bring the theory to agree with the data.

Let me also mention two theoretical contributions by Blazek \cite{bl}
and Yang \cite{ya2} who proposed new ways to analyse the multiparticle
data in small phase-space intervals. As I have no space to describe
their proposals in detail, I refer the reader directly to their written
versions published in this volume.

\section {Fluctuations at phase transition and event-by-event analysis}

The last topic discussed at the meeting concerned fluctuations occuring
at the phase transition. Antoniou presented results of the Athens group
\cite{ant} who took as the starting point the instanton model of the QCD
vacuum and looked at its behaviour close to the phase transition. They
found rather large rapidity fluctuations of the fractal type but only in
those events  which happen to satisfy the phase-transition
criteria. I find the problem important and thus I really look forward to
see the results of three-dimensional calculations promised by the
speaker.

Another important issue was stressed by Hwa \cite{hw2}. He pointed out the
essential difference between the determination of the fractal parameters
in case of dynamical systems and in case of systems of many particles.
In the dynamical system one can generate the {\it time sequence} and
thus estimate how fast the different trajectories diverge. In case of
multiparticle systems we do not have a time sequence and thus we have to
rely on {\it patterns}. The question in this case is: how different are
the patterns of different events. Hwa proposed to {\it measure} the
pattern of an event by the factorial moment associated with it. One can
then ask the question how this measure fluctuates from event to event.
Studying moments of this distribution provides a measure of
event-to-event fluctuation\footnote{To study the moments of the
factorial moments was suggested, in a  somewhat different context,  already some time ago \cite{bzd}.}. When
they are considered as function of bin size, it is possible to define
appropriate fractal dimensions which conveniently summarize the
information. For the details the reader is referred to the original
paper \cite{hw2}. I personally feel that this is an important conceptual
step in our thinking about the problem, although I am not fully
convinced that the proposed measure cannot be improved.

It thus clearly emerged from the work reported above that it is very
essential to be able to study the possible fractal behaviour in
event-by-event analysis. The feasibility of this program was
investigated in the paper presented by Ziaja \cite{zi2}. It was shown
that, in the framework of the $\alpha$-model, it is possible to improve
considerably the accuracy of the determination of the intermittency
parameters, as compared with the original proposal \cite{zib}. In the
discussion, it was pointed out by Eggers that the proposed corrections
must be tested on other models models before they can be considered
reliable.

Let me thus end by showing the only example of the event-by-event
analysis presented at the meeting \cite{se}: The distribution of the HBT
radii obtained from individual events by the NA49 collaboration is seen
in Fig.4. Although statistics is still limited (and the authors
themselves do not attach too much meaning to the details of the plot)
one clearly sees that the distribution is not symmetric, with a long tail
at large radii. I personally think that this is a hint of an interesting
phenomenon but one should obviously wait for more data before one starts
any theoretical speculations.

\vspace{8.5cm}

Fig.4. Distribution of HBT radii from event-by-event analysis of Pb-Pb
collisions \cite{se}.

\section {Conclusions}

A tentative summary of this summary can be formulated as follows.

(i) Studies of HBT correlations using the hydrodynamical model became an
effective tool for determining the space-time structure of particle
emission.

(ii) Fast progress is being made in Monte Carlo implementation of the
quantum interference, but controversies remain.

(iii) The problem of pion condensate is well understood and was even
analytically solved for Gaussian distributions.

(iv) An important step was achieved in theory of branching processes.

(v) Perturbative QCD works for global quantities but still fails to
describe local fluctuations.
 
(vi) Intermittency in $e^+e^-$ annihilation is confirmed by high-statistics data from OPAL.

(vii) Fluctuations at phase transition are being intensively studied.

(viii) Clear need for  event-by-event investigations emerges.

(ix) As expected, a new surprize from Bo arrived, and in time.

Let me close with apologies to all speakers whose work I have not been
able to report here  either for lack of space or (more often) for my
inability to summarize shortly their results.

\section* {Acknowledgements}
I would like to thank T.Csorgo for a kind hospitality at Matrahaza.
This work was supported in part by the KBN Grant No 2 P03B 086 14.


\section* {References}
\begin{thebibliography} {99}
\bibitem{eg}
H.Eggers, Phys.Rev.Lett. to be published and H.Eggers, these
proceedings.
\bibitem{an2}
B.Andersson, Lund report LUTP 97-36; 
B. Andersson et al. hep-ph/9807541, and these proceedings.
\bibitem{by}
G.Baym, Acta Phys. Pol. B29 (1998) 1839.
\bibitem{bk}
A.Bialas and A.Krzywicki, Phys. Letters B354 (1995) 134.
\bibitem{pr}
S.Pratt, Phys.Letters B301 (1993) 159.
\bibitem{zi}
T.Csorgo and J.Zimanyi, Phys.Rev.Letters 80 (1998) 916.
 J.Zimanyi, these proceedings and referrences quoted there.
\bibitem{zh}
H.Q.Zhang , these proceedings and references quoted there.
\bibitem{si}
Y.Sinyukov, Nucl.Phys. A566 (1994) 589c and these proceedings.
\bibitem{su}
N.Suzuki, these proceedings.
\bibitem{bz}
A.Bialas and K.Zalewski, hep-ph/9803408, Eur.Phys.J. C, in print; 
 hep-ph/9806435, Phys.Letters B, in print.
\bibitem{fu}
C.M.G.Lattes, Y.Fujimoto and S.Hasegawa, Phys.Rep. 65 (1980) 151.
\bibitem{blk}
J.-P.Blaizot and A.Krzywicki, Acta Phys. Pol. B27 (1996) 1687 and
references quoted there.
\bibitem{bj}
J.D.Bjorken, Acta Phys. Pol. B28 (1997) 2773 and references quoted there.
\bibitem{cse}
L.P.Csernai et al., these proceedings.
\bibitem{lo}
B.Lorstad, these proceedings.
\bibitem{is}
H.-Th. Elze and I.Sarcevic, Phys.Rev.Lett. 68 (1992) 1988.
\bibitem{elb}
H.Eggers, P.Lipa and B.Buschbeck, Phys.Rev.Lett. 79 (1997) 197.
\bibitem{fw}
K.Fialkowski and R.Wit, Z.Phys. C74 (1997) 145;
 Acta Phys. Pol. B28 (1997) 2039 and K.Fialkowski, these proceedings.
\bibitem{ar}
B.Andersson and M.Ringner, Nucl.Phys. B513 (1998) 627;
 Phys. Letters B421 (1998) 283; B.Andersson, Acta Phys. Pol. B29 (1998)
1885. 
 M.Ringner, these proceedings.
\bibitem{ah}
B.Andersson and W.Hoffman, Phys.Lett. B169 (1986) 364.
\bibitem{tn}
S.Todorova-Nova, these proceedings.
\bibitem{el}
J.Ellis and K.Geiger, Phys.Lett. B404 (1997) 230.
\bibitem{bu}
B.Buschbeck, these proceedings and private comm.
\bibitem{lon}
L.Lonnblad and T.Sjostrand, E.Phys.J. C2 (1998) 165 and L.Lonnblad,
these proceedings.
\bibitem{sm}
O.Smirnova, these proceedings.
\bibitem{we}
R.Weiner, these proceedings and references quoted there.
\bibitem{cs2}
T.Csorgo, and B.Lorstad, Heavy Ion Phys. 4 (1996) 221;
 T.Csorgo, these proceedings. 
\bibitem{ya}
F.Yano and S.Koonin, Phys.Lett. 78B (1978) 556; M.I.Podgoretskii,
Sov.J.Nucl.Phys. 37 (1983) 272.
\bibitem{se}
NA49 coll, Eur.Phys.J. C2 (1998) 359 and P.Seyboth, these proceedings.
\bibitem{bj2}
J.Bjorken, Phys. Rev. D45 (1992) 4077.
\bibitem{st}
A.Ster, these proceedings.
\bibitem{ha}
NA22 coll., Phys.Lett. B422 (1998) 359, and R.Hakobyan, these proceedings.
\bibitem{sh}
B.Schlei, these proceedings.
\bibitem{pr3}
S.Pratt, these proceedings.
\bibitem{le}
R.Lednicky, these proceedings.
\bibitem{ku}
V.I.Kuvshinov, these proceedings.
\bibitem{and}
I.V.Andreev, these proceedings.
\bibitem{cs3}
T.Csorgo et al., these proceedings.
\bibitem{gi}
A.Giovannini, R.Ugoccioni, these proceedings.
\bibitem{heg}
S.Hegyi, Phys.Lett. B411 (1997) 321; B417 (1998) 186; hep-ph/9709326 and 
 these proceedings.
\bibitem{pl}
R.Botet and M.Ploszajczak, Phys.Rev. E57 (1998) 7305; prepsint GANIL P 98 25
and M.Ploszajczak,  these proceedings.
\bibitem{phd}
Yu. Dokshitzer et al., Basics of Perturbative QCD, Ed.Frontieres (1991).
\bibitem{lup}
S.Lupia, hep-ph/9806493 and these proceedings
\bibitem{man}
D.Mangeol, these proceedings.
\bibitem{ki}
W.Kittel, these proceedings.
\bibitem{pqcd}
 W.Ochs and J.Wosiek, Phys.Lett. B289 (1992) 159;
Phys.Lett. B305 (1993) 144; Z.Phys. C68 (1995)269. Yu.L.Dokshitzer and
I.M.Dremin, Nucl.Phys. B402 (1993) 139. Ph.Brax, J.L.Meunier and
R.Peschanski, Z.Phys. C62 (1994) 649.
\bibitem{oc}
V.Khoze, S.Lupia and W.Ochs, Eur.Phys.J. C5 (1998) 77; 
W.Ochs, S.Lupia and J.Wosiek, hep-ph/9804419 and W.Ochs, these
proceedings.
\bibitem{hw}
R.Hwa, these proceedings and references quoted there.
\bibitem{fe}
R.C.Feynman, Photon-Hadron Interactions, Benjamin (1972).
\bibitem{bp}
A.Bialas and R.Peschanski, Nucl. Phys. B273 (1986) 703; Nucl.Phys. B308
(1988) 857.
\bibitem{ddk}
E.A.De Wolf, I.M.Dremin and W.Kittel, Phys.Rep. 270 (1996) 1.
\bibitem{che}
S.V.Chekanov, hep-ph/9806511 and these proceedings.
\bibitem{sar}
E.K.G. Sarkisyan, these proceedings.
\bibitem{bl}
M.Blazek, Int.J.Mod.Phys. A12 (1997) 839 and these proceedings.
\bibitem{ya2}
C.B.Yang, and X.Cai Phys.Rev. C58 (1998) 1183 and C.B.Yang, these
proceedings.
\bibitem{ant}
N.G.Antoniou, these proceedings and references quoted there.
\bibitem{hw2}
R.Hwa, Acta Phys.Pol. B27 (1996) 1789; Z.Cao and R.Hwa, nucl-th/9702015, and
R.Hwa, these proceedings.
\bibitem{bzd}
A.Bialas, A.Szczerba, K.Zalewski, Z.Phys. C46 (1990) 163. I.M. Dremin et al.,
Proc. XX Symp. Mult. Dynamics, Gut Holmecke (1990), p.459.
\bibitem{zi2}
R.Janik, B.Ziaja, hep-ph/9806227 and these proceedings.
\bibitem{zib}
A.Bialas and B.Ziaja, Phys.Letters B378 (1996) 319.

\end{thebibliography}
\end{document}